\documentclass{article}

\usepackage{arxiv}

\usepackage[utf8]{inputenc} % allow utf-8 input
\usepackage[T1]{fontenc}    % use 8-bit T1 fonts
\usepackage{hyperref}       % hyperlinks
\usepackage{url}            % simple URL typesetting
\usepackage{booktabs}       % professional-quality tables
\usepackage{amsfonts}       % blackboard math symbols
\usepackage{nicefrac}       % compact symbols for 1/2, etc.
\usepackage{microtype}      % microtypography
\usepackage{lipsum}
\usepackage{graphicx}
\graphicspath{ {./images/} }

\title{IllusionX: An LLM-powered mixed reality personal companion}

\author{
 Ramez Yousri \\
  Software Engineering\\
  Egyptian Chinese University\\
  Cairo, Egypt \\
  \texttt{ramez.yousri92@gmail.com} \\
  \And
 Zeyad Essam \\
  Software Engineering\\
  Egyptian Chinese University\\
  Cairo, Egypt \\
  \texttt{zessam268@gmail.com} \\
  \And
 Yehia Kareem \\
  Software Engineering\\
  Egyptian Chinese University\\
  Cairo, Egypt \\
  \texttt{yahya.karem55@gmail.com} \\
  \And
 Youstina Sherief \\
  Software Engineering\\
  Egyptian Chinese University\\
  Cairo, Egypt \\
  \texttt{youstinasherif21@gmail.com} \\
  \And
 Sherry Gamil \\
  Software Engineering\\
  Egyptian Chinese University\\
  Cairo, Egypt \\
  \texttt{sheerygamil@gmail.com} \\
  \And
 Soha Safwat \\
  Computer Science\\
  Egyptian Chinese University\\
  Cairo, Egypt \\
  \texttt{ssafwat@ecu.edu.eg} \\
}

\begin{document}
\maketitle
\begin{abstract}
Mixed Reality (MR) and Artificial Intelligence (AI) are increasingly becoming integral parts of our daily lives. Their applications range in fields from healthcare to education to entertainment. MR has opened a new frontier for such fields as well as new methods of enhancing user engagement. In this paper, We propose a new system one that combines the power of Large Language Models (LLMs) and mixed reality (MR) to provide a personalized companion for educational purposes. We present an overview of its structure and components as well tests to measure its performance. We found that our system is better in generating coherent information, however it's rather limited by the documents provided to it. This interdisciplinary approach aims to provide a better user experience and enhance user engagement. The user can interact with the system through a custom-design smart watch, smart glasses and a mobile app. 
\end{abstract}


\section{Introduction}
Affective computing\cite{cabada2023affective} aims to improve user interaction with computers and machines. It does so by capturing the user's emotions and psychological states. The main tasks of affective computing systems include emotion recognition, facial expression, and engagement detection. This field of computing is related to Human-Computer Interaction (HCI) or more broadly to Human-Machine Interaction (HMI). To achieve its intended tasks, an affective computing system captures the emotions and affective states of the user through verbal and non-verbal communication signals. This means that usually machine learning algorithms are used to determine the state, and then the system decides the appropriate response that it should give the user based on the model's output. Examples of such algorithms include Support Vector Machines (SVMs) which were used to detect user's emotions through speech \cite{royal2023new}. Convolutional Neural Networks (CNNs) \cite{sarvakar2023facial} by facial recognition. CNNs and LSTMs can also detect emotions by analyzing EEG patterns as in \cite{iyer2023cnn}.

Education is one of the most important aspects of human life. The rise of new advanced computing technologies like artificial intelligence and affective computing made personalized and adaptive learning possible. In recent years, AI has become a boom in the field of education research, it has been used to improve the quality of education as well as provide a source for personal learning as exemplified in \cite{halaweh2023chatgpt}, which discusses strategies of incorporating ChatGPT in education.

Another field that promises a huge opportunity to improve education is Mixed Reality (MR), which encompasses Virtual Reality (VR) and Augmented Reality (AR). MR has proven to be a strong candidate for a better learning experience, especially adding the ability of online learning or e-learning since the COVID-19 pandemic.

What we present in this paper, is a new system that combines MR, AI, and affective computing to provide a personal companion that aids the user in learning, as well as aids the instructor in preparing interactive lessons and curricula. In section 2, we summarize the most important concepts in Large Language Models (LLMs), MR systems, and their use in education. We also present the pros and cons of each in education. Section 3, we present the system specification, in terms of functional and non-functional requirements and functionalities of the system. In Section 4, we talk about the system design and components in more detail. In Section 5, we present the surveys we conducted as well as some tests to test the efficiency of our system in achieving its intended functionalities. We then talk about potential challenges and future directions in section 6. We then conclude the paper with a summary of the system and its functionalities. As per the current knowledge of the authors, this is the first system that combines LLMs, MR, and affective computing to provide a comprehensive system tailored to the educational sector.

\section{Literature Review}
\subsection{Large Language Models}
During the past few years, Large Language models made huge advances in natural language processing (NLP), surpassing the capabilities of more traditional approaches to NLP, like Neural Language Models (NLMs). This is due to two fundamental changes in their algorithm, first, they are dependent on the idea of pre-training before the actual training of the model, and their architecture is based on Transformers. There are 3 main types of LLMs, and these are:
\begin{itemize}
	\item \textit{Autoregressive LLMs (Decoder only)}: The most prevalent and widely employed model for LLMs. This type of LLMs can be seen in GPT-like models such as LLaMa2 \cite{touvron2023llama}, and BLOOM \cite{workshop2022bloom}. In this framework, each generated token relies on the tokens produced before it. Tokens are generated sequentially, one at a time. During the training phase, the decoder is provided with the specific next token, while in the inference phase, the decoder independently generates tokens based on its acquired knowledge.
	\item \textit{Autoencoding LLMs (Encoder only)}: In contrast to decoder-only models, auto-encoding models, also known as encoder-only models, employ an encoder to produce text. These models are alternatively referred to as masked language models (MLMs). MLMs operate bidirectionally, predicting the current word (token) based on all words (tokens) in the sequence, both preceding and following the current word. In the training process, words designated for generation are randomly chosen and masked (indicated by a special token [MASK] or replaced with another random token). This compels the model to capture bidirectional context while making predictions. Prominent examples of such models include BERT\cite{devlin2018bert}.
	\item \textit{Encoder-Decoder LLMs (Seq-to-Seq)}: These models are designed to generate sequences based on another given sequence as input, producing a sequence y1, y2..., yn given a sequence x1, x2...., xn. Consequently, they are also referred to as sequence-to-sequence models. In the training phase, a corrupted sequence is presented to the model, and it endeavors to predict the correct, uncorrupted sequence. Various sequence corruption methods are employed, such as document rotation, sentence permutation, and token deletion or masking. A notable example of this type is T5 \cite{raffel2020exploring}.

\end{itemize}
LLMs surpassed their predecessors in various tasks including question answering and text summarization which proves them an excellent candidate to use in education.
\subsubsection{Their use in education}
The powerful capabilities of LLMs led them to become a candidate for their use in education, we provide a simple overview of examples of their use cases in education. Question generation is one of the most fundamental tasks in education or learning in general. A robust QG system could provide its users with a chance to test their knowledge as well as understand the material more deeply. In\cite{elkins2023useful}, the authors showed that the questions generated by LLMs are high quality and sufficiently useful. This shows a great promise for the integration of LLMs in the classroom. In \cite{caines2023application}, showed that larger language models have shown better text generation capabilities, than smaller ones. They also recommend careful prompting and post-generation editing of the generated text before use. The authors of \cite{dai2023can}, showed that ChatGPT provides more detailed that fluently and coherently
summarizes students’ performance than human instructors. It can also provide feedback on the process of students completing the
task, which benefits students developing learning skills. ChatGPT also can be used to help adults with self-learning goals and tasks \cite{lin2023exploring}. Large Language Models have been used in healthcare education\cite{benitez2024harnessing,sallam2023utility,abd2023large}, engineering education\cite{fan2023large,lesage2024exploring} and language learning\cite{caines2023application}.

\subsubsection{Challenges}
Even though LLMs present huge advantages and advancements, they also present challenges. These challenges include: the huge amount of training data needed \cite{chen2023seed,hoffmann2022training}, ethical concerns\cite{sennrich2015neural}, adversarial attacks\cite{liu2020adversarial,shayegani2023survey},hallucination\cite{ye2023cognitive}. For a more detailed discussion of the challenges, you can check \cite{hadi2023large}.

\textbf{Summary:} Large Language Models show huge promise when it comes to education-related tasks like question generation, and content generation. They have also been used in engineering, healthcare, and language learning. This shows that it can as a compliment to traditional learning methods.
\subsection{Mixed Reality Systems}
Human-Computer Interaction (HCI) is one important field of research to improve interaction with computing systems. Researchers have recently found that user interactivity can be improved by incorporating visualizing strategies to immerse the user in virtual environments. To achieve this, various ideas were proposed including Virtual Reality (VR)\cite{anthes2016state,biocca1992virtual}, Augmented Reality (AR)\cite{dargan2023augmented}, and Mixed Reality (MR)\cite{rokhsaritalemi2020review}. MR aims to maximize user interaction with the system or environment compared to its counterparts. It's rather difficult to define what exactly is MR, however in \cite{speicher2019mixed}, the authors interviewed experts to put a definition to the term MR. They also came up with a framework to explain what an MR system is. They came up with the following components or dimensions of defining an MR system: number of physical and virtual environments, number of users, level of immersion, level of virtuality, and degree of interaction.
\subsubsection{Their use in education}
The abilities provided by MR systems led researchers to address their use in education. In \cite{aguayo2023using,cho2023designing}, the authors discussed using an MR system in environmental education. The purpose of environmental education is to increase awareness of climate change and global warming and its effects. In \cite{acampora2023serious}, an MR-based game was designed to help police officers in crime scene tagging. Roadmaps for integrating extended and MR in engineering education have been developed for example \cite{carberry2023roadmap}. In this paper, the authors developed a roadmap for designing tools that would help chemical engineers while learning.
\subsubsection{Challenges}
While the potential to use MR systems in education has proven quite effective, it has its downsides as well. One of these is that it can overwhelm the user with information, as the user would have to process a lot of input signals provided to him by both the virtual and physical environments. Some pedagogical challenges include the need to keep the information presented to a minimum, and the need to ensure that the learning activities include well-designed interaction with the peers within the classroom. Some of the technical challenges for MR include: most of the devices are still in their infancy, limited battery life, and uncomfortable feelings for users when such devices as headsets are worn for a long time. MR systems also might cause disorientation when used while doing something that needs focus like driving or even walking\cite{xue2019review}.  One of the biggest challenges that needs to be addressed is developing meaningful MR environments for learning.

\textbf{Summary}: The integration of Mixed Reality (MR) systems, including Virtual Reality (VR) and Augmented Reality (AR), in education enhances user interaction and creates immersive learning experiences. Defined by factors like environments, users, immersion, virtuality, and interaction, MR finds applications in environmental education, police training, and engineering education. However, challenges like information overload, pedagogical considerations, and technical limitations, such as device infancy and discomfort during prolonged use, need addressing. Striking a balance between content delivery and resolving these challenges is crucial for realizing the full potential of MR systems in education.

\section{System Specification}
Our LLM-powered MR-based system aims to provide a better user experience as well as a more personalized companion for educational purposes for its users. The system consists of two main components with more sub-components mentioned in section 4. The two main components are a software application (AI, backend, and mobile app) and hardware devices (smart glasses and a smartwatch). In this section we detail the scope as well as the requirements for the system:
\begin{figure}[!ht]
	\centering
	\includegraphics[scale=0.3]{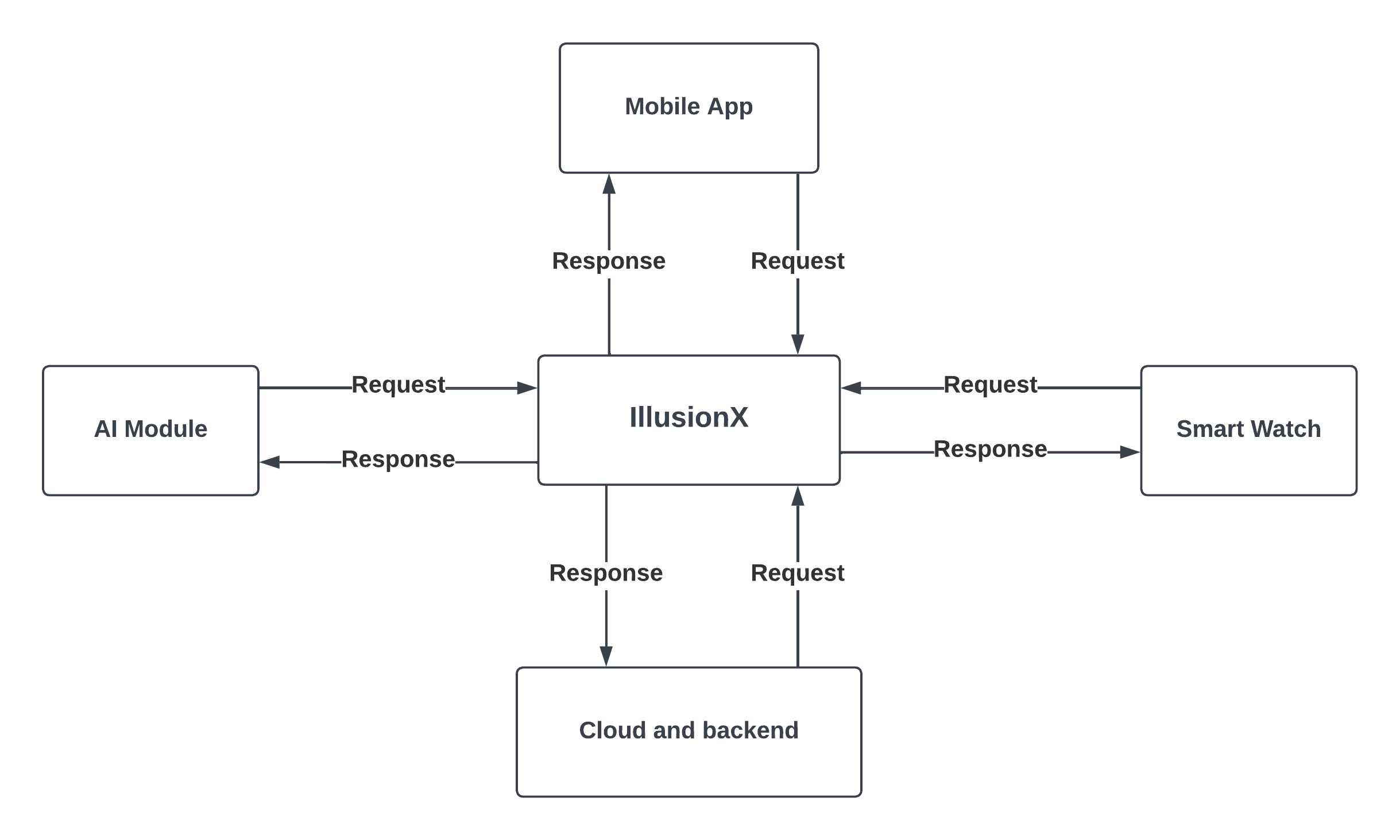}
	\caption{IllusionX Context Diagram}
	\label{fig:enter-label}
\end{figure}
\subsection{System Functionalities}
The project aims to combine LLMs as well as MR technology to provide the following functionalities:
\begin{itemize}
	\item \textit{Information Retrieval}: The companion will be able to provide information on a wide variety of topics when prompted by the user.
	\item \textit{Educational capabilities}: It will have the ability to teach individuals or explain complex topics in various levels of understanding to them, serving as an educational resource.
	\item \textit{Task assistance}: The system should be able to help users with their learning-related tasks like organizing notes or summarizing certain texts
	\item \textit{Conversational and immersive interface:} Users will be able to engage in casual conversations with the companion, treating it as a friendly and interactive virtual friend.
\end{itemize}
\subsection{Functional Requirements}
The following are the functional requirements of the system:
\begin{itemize}
	\item \textit{RE1}: The system should be able to retrieve and provide accurate information on various topics when prompted by the user.
	\item \textit{RE2}: The system should be able to provide a conversational and immersive experience to the user.
	\item \textit{RE3}: The system should be able to let the users create their custom chatbots based on their needs.
	\item \textit{RE4}:  The system should provide multiple ways to authenticate users.
	\item \textit{RE5}: The system should be able to accessible to its users 95\% of the time.
\end{itemize}
\subsection{Non-Functional Requirements}
The following are the non-functional requirements of the system:
\begin{itemize}
	\item \textit{N-RE1}: The system should have an intuitive and user-friendly interface.
	\item \textit{N-RE2}: The system should be able to handle a large number of concurrent users without affecting the response time, it should also be scalable.
	\item \textit{N-RE3}: The system should contain robust security measures to ensure data protection and privacy.
\end{itemize}
\subsection{Ethical Considerations}
Due to the use of LLMs in our systems, some ethical considerations should be mentioned. The first one is hallucination, the generalization capabilities of LLMs could give rise to certain informational inaccuracies when it comes to knowledge generation \cite{yu2022survey}. Hallucination is when an LLM generates knowledge that's not grounded in truth or completely wrong, this is a critical problem when it comes to education. To combat such a problem, many approaches have been devised including parameter adaptation like \cite{wang2023easyedit,lee2022factuality,sun2023contrastive}, leveraging external knowledge like \cite{wang2023knowledgpt,borgeaud2022improving,trivedi2022interleaving,liu2023reta} and assessment feedback like \cite{stiennon2020learning,agrawal2023language}. We aim to reduce hallucination by leveraging external knowledge through knowledge embedding.

\section{System Design and Components}
\subsection{LLM}
The first component in our system is the LLM. Due to the high costs and time needed for training a fully-fledged LLM, we ended up using a pre-trained LLM through an API. The available options were ChatGPT\cite{halaweh2023chatgpt}, PaLM2\cite{anil2023palm}, and Google Gemini\cite{saeidnia2023welcome,team2023gemini}. We chose PaLM2 due to its availability, free of cost, and ease of use. PaLM2 API can be accessed using Python, which makes it a good option for our use as we use the Python programming language in our backend and API.
\subsection{API}
Another integral component is the IllusionX API, a platform developed with FastAPI\cite{lathkar2023getting,lathkar2023introduction} and PostgreSQL\cite{makris2021mongodb}. FastAPI, known for its high performance and simplicity, provides the backbone of our API, ensuring lightning-fast responses and seamless scalability to meet the scalability requirement of our system. Our choice of PostgreSQL came due to the paper \cite{makris2021mongodb} in which the authors showed the superiority of PostgreSQL over Mongodb in various business scenarios. We also use Alembic as a database migration tool, and Pydantic for schema validation.
\subsection{Mobile App}
We also made our system accessible through a cross-platform mobile application developed in Flutter. The app contains the following functionality: Login, Signup, chat, and also controlling available bots (agents). Agents are specialized chatbots for different fields. It has a user-friendly interface to facilitate its use and adoption by the target audience.
\subsection{Smart Glasses and Smart Watch}
The hardware components of our system are a smart watch and smart glasses. Smart glasses use AR displays \cite{dargan2023augmented} to overlay certain digital information on the lenses\cite{koutromanos2023augmented,england2023comparison}. The smartwatch, however, can act as a separate hardware piece. The smartwatch is made up of a custom design System-on-Chip (SoC)\cite{chuah2016wearable} to incorporate the ability to generate audio and visual responses based on the user's request. The watch combined with the glasses could immerse the user in a virtual environment combined with the users' environment, providing an immersive and interactive MR environment. The detailed design, testing, and verification of the hardware components will be published in a later paper.

\section{Testing and Results}
\subsection{Technology adoption and acceptance}
To assess the adoption of our system by our target audience, we conducted a survey and sent it to representatives of our target users. We found that around 87.5\% of the users who filled out the survey are interested in personal companions helping them in their learning and daily tasks. 67\% are interested in the system containing both text and voice commands. 62.5\% of those who answered stated that they would use the system for information retrieval, thus making it the most requested capability of the system. The rest of the functions chosen by the users as well as the percentage of each one is described in table \ref{suvey}.
\begin{table}[!ht]
	\centering
	\begin{tabular}{|c|c|}
	\hline
    	\textbf{Functionality} & \textbf{Percentage (\%)}  \\
    	\hline
    	Information retrieval & 62.5\\\hline
    	Task management & 52.5\\\hline
    	Social Interaction & 51.2\\
    	\hline
   	 
	\end{tabular}
	\caption{Requested functionalities of the system. NB: the user could choose multiple functionalities per response}
	\label{suvey}

\end{table}
\subsection{LLM testing for educational tasks}
We tested our system on 3 essential tasks for learning, these include generating course description and outline, lesson generation, and Q\&A. We also compared our system with PaLM. Even though we used the PaLM model as our foundation model, we tested our knowledge embedding module to see if it improves the extracted results over the vanilla one. We embed knowledge by providing PDF documents to the model.
\begin{itemize}
	\item \textit{Course descriptions}: In generating course descriptions and outlines, we tried out 3 different prompts with the vanilla PaLM. We tested it in 2 fields, Artificial Intelligence (AI) and nanoelectronics. We found out that the vanilla PaLM generated coherent curriculums but way too broad for an engineering student or too non-technical. We tested our system on the same topics, we found that even though our system is limited by the given document or knowledge embedded within the prompt, it generated more detailed course descriptions.
	\item \textit{Lesson Generation}: When it comes to lesson generation, both our system and the vanilla PaLM need careful prompting to generate coherent lessons. However, we were able to marginally reduce hallucination but our system is limited to the documents provided.
	\item \textit{Q\&A}: Vanilla PaLM can sometimes hallucinate or explain concepts that aren't what the user asked for. Our system, however, can answer more technical and detailed questions. It's still limited by the information provided within the documents.
\end{itemize}
A summary of our system's advantages and disadvantages:
\begin{table}[!ht]
	\centering
	\begin{tabular}{|p{30mm}|p{40mm}|}
	\hline
     	\textbf{Advantages}& \textbf{Disadvantages}  \\\hline
     	More coherent lesson descriptions and course outlines& Limited by the provided documents\\\hline
     	Offers more technical Q\&A& It can hallucinate the information within the document, however, it's still close to being correct\\\hline
	\end{tabular}
	\caption{Advantages and Disadvantages of IllusionX}
	\label{tab:my_label}
\end{table}

\textbf{Guidelines for effective prompts}:
Based on our tests, we found the following prompt guidelines help in generating more effective responses:
\begin{enumerate}
	\item Make sure to give a role to the model, e.g., acting as a university professor or, you are a university professor.
	\item Make sure to give as much detail as possible regarding the information on the lesson or course without the need to provide the technical aspects of it.
	\item Make sure your provided documents are relevant to the topic and organized clearly.
\end{enumerate}
Example prompt: \textit{Acting as a university professor, generate a detailed course description and outline for an introductory course on VLSI design. The course should target junior engineering students. The course should cover the basics of VLSI as well as the process of designing VLSI systems and manufacturing. The course spans 12 weeks.}

\section{Future Directions}
While our system shows marginal improvement in education-based goals and tasks, there are yet improvements to be made to the system. One such improvement would be to try to reduce the hallucination of our LLM, whether that be through adding special, custom-designed components or creating our LLM from scratch. Another avenue of exploration would be incorporating features for those with special needs to help them benefit from the capabilities of the system. It also could be a multi-lingual, multi-modal model which can improve the overall effectiveness of the system. Hardware has a huge room for improvement in terms of design and technology as well.

\section{Conclusion}
In this paper, we presented a new LLM-powered mixed reality system to use in education. We showed the various components of the system as well as provided results of the tests we provided to assess technology adoption and its performance in its intended tasks. We also discussed possible limitations and ethical considerations related to the system and mentioned possible future directions for its improvement.

\bibliography{references}

\begin{thebibliography}{10}

\bibitem{cabada2023affective}
Ram{\'o}n~Zatarain Cabada, H{\'e}ctor Manuel~C{\'a}rdenas L{\'o}pez, and Hugo~Jair Escalante.
\newblock Affective computing.
\newblock In {\em Multimodal Affective Computing: Technologies and Applications in Learning Environments}, pages 3--20. Springer, 2023.

\bibitem{royal2023new}
G~Jayasree Royal and L~Rama Parvathy.
\newblock New speech emotion recognition analysis to increase accuracy using svm over decision tree.
\newblock {\em Journal of Survey in Fisheries Sciences}, 10(1S):1943--1951, 2023.

\bibitem{sarvakar2023facial}
Ketan Sarvakar, R~Senkamalavalli, S~Raghavendra, J~Santosh Kumar, R~Manjunath, and Sushma Jaiswal.
\newblock Facial emotion recognition using convolutional neural networks.
\newblock {\em Materials Today: Proceedings}, 80:3560--3564, 2023.

\bibitem{iyer2023cnn}
Abhishek Iyer, Srimit~Sritik Das, Reva Teotia, Shishir Maheshwari, and Rishi~Raj Sharma.
\newblock Cnn and lstm based ensemble learning for human emotion recognition using eeg recordings.
\newblock {\em Multimedia Tools and Applications}, 82(4):4883--4896, 2023.

\bibitem{halaweh2023chatgpt}
Mohanad Halaweh.
\newblock Chatgpt in education: Strategies for responsible implementation.
\newblock 2023.

\bibitem{touvron2023llama}
Hugo Touvron, Louis Martin, Kevin Stone, Peter Albert, Amjad Almahairi, Yasmine Babaei, Nikolay Bashlykov, Soumya Batra, Prajjwal Bhargava, Shruti Bhosale, et~al.
\newblock Llama 2: Open foundation and fine-tuned chat models.
\newblock {\em arXiv preprint arXiv:2307.09288}, 2023.

\bibitem{workshop2022bloom}
BigScience Workshop, Teven~Le Scao, Angela Fan, Christopher Akiki, Ellie Pavlick, Suzana Ili{\'c}, Daniel Hesslow, Roman Castagn{\'e}, Alexandra~Sasha Luccioni, Fran{\c{c}}ois Yvon, et~al.
\newblock Bloom: A 176b-parameter open-access multilingual language model.
\newblock {\em arXiv preprint arXiv:2211.05100}, 2022.

\bibitem{devlin2018bert}
Jacob Devlin, Ming-Wei Chang, Kenton Lee, and Kristina Toutanova.
\newblock Bert: Pre-training of deep bidirectional transformers for language understanding.
\newblock {\em arXiv preprint arXiv:1810.04805}, 2018.

\bibitem{raffel2020exploring}
Colin Raffel, Noam Shazeer, Adam Roberts, Katherine Lee, Sharan Narang, Michael Matena, Yanqi Zhou, Wei Li, and Peter~J Liu.
\newblock Exploring the limits of transfer learning with a unified text-to-text transformer.
\newblock {\em The Journal of Machine Learning Research}, 21(1):5485--5551, 2020.

\bibitem{elkins2023useful}
Sabina Elkins, Ekaterina Kochmar, Iulian Serban, and Jackie~CK Cheung.
\newblock How useful are educational questions generated by large language models?
\newblock In {\em International Conference on Artificial Intelligence in Education}, pages 536--542. Springer, 2023.

\bibitem{caines2023application}
Andrew Caines, Luca Benedetto, Shiva Taslimipoor, Christopher Davis, Yuan Gao, Oeistein Andersen, Zheng Yuan, Mark Elliott, Russell Moore, Christopher Bryant, et~al.
\newblock On the application of large language models for language teaching and assessment technology.
\newblock {\em arXiv preprint arXiv:2307.08393}, 2023.

\bibitem{dai2023can}
Wei Dai, Jionghao Lin, Hua Jin, Tongguang Li, Yi-Shan Tsai, Dragan Ga{\v{s}}evi{\'c}, and Guanliang Chen.
\newblock Can large language models provide feedback to students? a case study on chatgpt.
\newblock In {\em 2023 IEEE International Conference on Advanced Learning Technologies (ICALT)}, pages 323--325. IEEE, 2023.

\bibitem{lin2023exploring}
Xi~Lin.
\newblock Exploring the role of chatgpt as a facilitator for motivating self-directed learning among adult learners.
\newblock {\em Adult Learning}, page 10451595231184928, 2023.

\bibitem{benitez2024harnessing}
Trista~M Ben{\'\i}tez, Yueyuan Xu, J~Donald Boudreau, Alfred Wei~Chieh Kow, Fernando Bello, Le~Van Phuoc, Xiaofei Wang, Xiaodong Sun, Gilberto Ka-Kit Leung, Yanyan Lan, et~al.
\newblock Harnessing the potential of large language models in medical education: promise and pitfalls.
\newblock {\em Journal of the American Medical Informatics Association}, page ocad252, 2024.

\bibitem{sallam2023utility}
Malik Sallam.
\newblock The utility of chatgpt as an example of large language models in healthcare education, research and practice: Systematic review on the future perspectives and potential limitations.
\newblock {\em medRxiv}, pages 2023--02, 2023.

\bibitem{abd2023large}
Alaa Abd-Alrazaq, Rawan AlSaad, Dari Alhuwail, Arfan Ahmed, Padraig~Mark Healy, Syed Latifi, Sarah Aziz, Rafat Damseh, Sadam~Alabed Alrazak, Javaid Sheikh, et~al.
\newblock Large language models in medical education: Opportunities, challenges, and future directions.
\newblock {\em JMIR Medical Education}, 9(1):e48291, 2023.

\bibitem{fan2023large}
Angela Fan, Beliz Gokkaya, Mark Harman, Mitya Lyubarskiy, Shubho Sengupta, Shin Yoo, and Jie~M Zhang.
\newblock Large language models for software engineering: Survey and open problems.
\newblock {\em arXiv preprint arXiv:2310.03533}, 2023.

\bibitem{lesage2024exploring}
Jonathan Lesage, Robert Brennan, Sarah~Elaine Eaton, Beatriz Moya, Brenda McDermott, Jason Wiens, and Kai Herrero.
\newblock Exploring natural language processing in mechanical engineering education: Implications for academic integrity.
\newblock {\em International Journal of Mechanical Engineering Education}, 52(1):88--105, 2024.

\bibitem{chen2023seed}
Zui CHen, Lei Cao, Sam Madden, Ju~Fan, Nan Tang, Zihui Gu, Zeyuan Shang, Chunwei Liu, Michael Cafarella, and Tim Kraska.
\newblock Seed: Simple, efficient, and effective data management via large language models.
\newblock {\em arXiv preprint arXiv:2310.00749}, 2023.

\bibitem{hoffmann2022training}
Jordan Hoffmann, Sebastian Borgeaud, Arthur Mensch, Elena Buchatskaya, Trevor Cai, Eliza Rutherford, Diego de~Las Casas, Lisa~Anne Hendricks, Johannes Welbl, Aidan Clark, et~al.
\newblock Training compute-optimal large language models.
\newblock {\em arXiv preprint arXiv:2203.15556}, 2022.

\bibitem{sennrich2015neural}
Rico Sennrich, Barry Haddow, and Alexandra Birch.
\newblock Neural machine translation of rare words with subword units.
\newblock {\em arXiv preprint arXiv:1508.07909}, 2015.

\bibitem{liu2020adversarial}
Xiaodong Liu, Hao Cheng, Pengcheng He, Weizhu Chen, Yu~Wang, Hoifung Poon, and Jianfeng Gao.
\newblock Adversarial training for large neural language models.
\newblock {\em arXiv preprint arXiv:2004.08994}, 2020.

\bibitem{shayegani2023survey}
Erfan Shayegani, Md~Abdullah~Al Mamun, Yu~Fu, Pedram Zaree, Yue Dong, and Nael Abu-Ghazaleh.
\newblock Survey of vulnerabilities in large language models revealed by adversarial attacks.
\newblock {\em arXiv preprint arXiv:2310.10844}, 2023.

\bibitem{ye2023cognitive}
Hongbin Ye, Tong Liu, Aijia Zhang, Wei Hua, and Weiqiang Jia.
\newblock Cognitive mirage: A review of hallucinations in large language models.
\newblock {\em arXiv preprint arXiv:2309.06794}, 2023.

\bibitem{hadi2023large}
Muhammad~Usman Hadi, Rizwan Qureshi, Abbas Shah, Muhammad Irfan, Anas Zafar, Muhammad~Bilal Shaikh, Naveed Akhtar, Jia Wu, Seyedali Mirjalili, et~al.
\newblock Large language models: a comprehensive survey of its applications, challenges, limitations, and future prospects.
\newblock {\em Authorea Preprints}, 2023.

\bibitem{anthes2016state}
Christoph Anthes, Rub{\'e}n~Jes{\'u}s Garc{\'\i}a-Hern{\'a}ndez, Markus Wiedemann, and Dieter Kranzlm{\"u}ller.
\newblock State of the art of virtual reality technology.
\newblock In {\em 2016 IEEE aerospace conference}, pages 1--19. IEEE, 2016.

\bibitem{biocca1992virtual}
Frank Biocca.
\newblock Virtual reality technology: A tutorial.
\newblock {\em Journal of communication}, 42(4):23--72, 1992.

\bibitem{dargan2023augmented}
Shaveta Dargan, Shally Bansal, Munish Kumar, Ajay Mittal, and Krishan Kumar.
\newblock Augmented reality: A comprehensive review.
\newblock {\em Archives of Computational Methods in Engineering}, 30(2):1057--1080, 2023.

\bibitem{rokhsaritalemi2020review}
Somaiieh Rokhsaritalemi, Abolghasem Sadeghi-Niaraki, and Soo-Mi Choi.
\newblock A review on mixed reality: Current trends, challenges and prospects.
\newblock {\em Applied Sciences}, 10(2):636, 2020.

\bibitem{speicher2019mixed}
Maximilian Speicher, Brian~D Hall, and Michael Nebeling.
\newblock What is mixed reality?
\newblock In {\em Proceedings of the 2019 CHI conference on human factors in computing systems}, pages 1--15, 2019.

\bibitem{aguayo2023using}
Claudio Aguayo and Chris Eames.
\newblock Using mixed reality (xr) immersive learning to enhance environmental education.
\newblock {\em The Journal of Environmental Education}, 54(1):58--71, 2023.

\bibitem{cho2023designing}
Yongjoo Cho and Kyoung~Shin Park.
\newblock Designing immersive virtual reality simulation for environmental science education.
\newblock {\em Electronics}, 12(2):315, 2023.

\bibitem{acampora2023serious}
Giovanni Acampora, Pasquale Trinchese, Roberto Trinchese, and Autilia Vitiello.
\newblock A serious mixed-reality game for training police officers in tagging crime scenes.
\newblock {\em Applied Sciences}, 13(2):1177, 2023.

\bibitem{carberry2023roadmap}
Deborah~E Carberry, Khosrow Bagherpour, Christian Beenfeldt, John~M Woodley, Seyed~Soheil Mansouri, and Martin~P Andersson.
\newblock A roadmap for designing extended reality tools to teach unit operations in chemical engineering: Learning theories \& shifting pedagogies.
\newblock {\em Digital Chemical Engineering}, 6:100074, 2023.

\bibitem{xue2019review}
Yukang Xue.
\newblock A review on intelligent wearables: Uses and risks.
\newblock {\em Human Behavior and Emerging Technologies}, 1(4):287--294, 2019.

\bibitem{yu2022survey}
Wenhao Yu, Chenguang Zhu, Zaitang Li, Zhiting Hu, Qingyun Wang, Heng Ji, and Meng Jiang.
\newblock A survey of knowledge-enhanced text generation.
\newblock {\em ACM Computing Surveys}, 54(11s):1--38, 2022.

\bibitem{wang2023easyedit}
Peng Wang, Ningyu Zhang, Xin Xie, Yunzhi Yao, Bozhong Tian, Mengru Wang, Zekun Xi, Siyuan Cheng, Kangwei Liu, Guozhou Zheng, et~al.
\newblock Easyedit: An easy-to-use knowledge editing framework for large language models.
\newblock {\em arXiv preprint arXiv:2308.07269}, 2023.

\bibitem{lee2022factuality}
Nayeon Lee, Wei Ping, Peng Xu, Mostofa Patwary, Pascale~N Fung, Mohammad Shoeybi, and Bryan Catanzaro.
\newblock Factuality enhanced language models for open-ended text generation.
\newblock {\em Advances in Neural Information Processing Systems}, 35:34586--34599, 2022.

\bibitem{sun2023contrastive}
Weiwei Sun, Zhengliang Shi, Shen Gao, Pengjie Ren, Maarten de~Rijke, and Zhaochun Ren.
\newblock Contrastive learning reduces hallucination in conversations.
\newblock In {\em Proceedings of the AAAI Conference on Artificial Intelligence}, volume~37, pages 13618--13626, 2023.

\bibitem{wang2023knowledgpt}
Xintao Wang, Qianwen Yang, Yongting Qiu, Jiaqing Liang, Qianyu He, Zhouhong Gu, Yanghua Xiao, and Wei Wang.
\newblock Knowledgpt: Enhancing large language models with retrieval and storage access on knowledge bases.
\newblock {\em arXiv preprint arXiv:2308.11761}, 2023.

\bibitem{borgeaud2022improving}
Sebastian Borgeaud, Arthur Mensch, Jordan Hoffmann, Trevor Cai, Eliza Rutherford, Katie Millican, George~Bm Van Den~Driessche, Jean-Baptiste Lespiau, Bogdan Damoc, Aidan Clark, et~al.
\newblock Improving language models by retrieving from trillions of tokens.
\newblock In {\em International conference on machine learning}, pages 2206--2240. PMLR, 2022.

\bibitem{trivedi2022interleaving}
Harsh Trivedi, Niranjan Balasubramanian, Tushar Khot, and Ashish Sabharwal.
\newblock Interleaving retrieval with chain-of-thought reasoning for knowledge-intensive multi-step questions.
\newblock {\em arXiv preprint arXiv:2212.10509}, 2022.

\bibitem{liu2023reta}
Jiongnan Liu, Jiajie Jin, Zihan Wang, Jiehan Cheng, Zhicheng Dou, and Ji-Rong Wen.
\newblock Reta-llm: A retrieval-augmented large language model toolkit.
\newblock {\em arXiv preprint arXiv:2306.05212}, 2023.

\bibitem{stiennon2020learning}
Nisan Stiennon, Long Ouyang, Jeffrey Wu, Daniel Ziegler, Ryan Lowe, Chelsea Voss, Alec Radford, Dario Amodei, and Paul~F Christiano.
\newblock Learning to summarize with human feedback.
\newblock {\em Advances in Neural Information Processing Systems}, 33:3008--3021, 2020.

\bibitem{agrawal2023language}
Ayush Agrawal, Lester Mackey, and Adam~Tauman Kalai.
\newblock Do language models know when they're hallucinating references?
\newblock {\em arXiv preprint arXiv:2305.18248}, 2023.

\bibitem{anil2023palm}
Rohan Anil, Andrew~M Dai, Orhan Firat, Melvin Johnson, Dmitry Lepikhin, Alexandre Passos, Siamak Shakeri, Emanuel Taropa, Paige Bailey, Zhifeng Chen, et~al.
\newblock Palm 2 technical report.
\newblock {\em arXiv preprint arXiv:2305.10403}, 2023.

\bibitem{saeidnia2023welcome}
Hamid~Reza Saeidnia.
\newblock Welcome to the gemini era: Google deepmind and the information industry.
\newblock {\em Library Hi Tech News}, 2023.

\bibitem{team2023gemini}
Gemini Team, Rohan Anil, Sebastian Borgeaud, Yonghui Wu, Jean-Baptiste Alayrac, Jiahui Yu, Radu Soricut, Johan Schalkwyk, Andrew~M Dai, Anja Hauth, et~al.
\newblock Gemini: a family of highly capable multimodal models.
\newblock {\em arXiv preprint arXiv:2312.11805}, 2023.

\bibitem{lathkar2023getting}
Malhar Lathkar.
\newblock Getting started with fastapi.
\newblock In {\em High-Performance Web Apps with FastAPI: The Asynchronous Web Framework Based on Modern Python}, pages 29--64. Springer, 2023.

\bibitem{lathkar2023introduction}
Malhar Lathkar.
\newblock Introduction to fastapi.
\newblock In {\em High-Performance Web Apps with FastAPI: The Asynchronous Web Framework Based on Modern Python}, pages 1--28. Springer, 2023.

\bibitem{makris2021mongodb}
Antonios Makris, Konstantinos Tserpes, Giannis Spiliopoulos, Dimitrios Zissis, and Dimosthenis Anagnostopoulos.
\newblock Mongodb vs postgresql: A comparative study on performance aspects.
\newblock {\em GeoInformatica}, 25:243--268, 2021.

\bibitem{koutromanos2023augmented}
George Koutromanos and Georgia Kazakou.
\newblock Augmented reality smart glasses use and acceptance: A literature review.
\newblock {\em Computers \& Education: X Reality}, 2:100028, 2023.

\bibitem{england2023comparison}
A~England, J~Thompson, S~Dorey, S~Al-Islam, M~Long, C~Maiorino, and MF~McEntee.
\newblock A comparison of perceived image quality between computer display monitors and augmented reality smart glasses.
\newblock {\em Radiography}, 29(3):641--646, 2023.

\bibitem{chuah2016wearable}
Stephanie Hui-Wen Chuah, Philipp~A Rauschnabel, Nina Krey, Bang Nguyen, Thurasamy Ramayah, and Shwetak Lade.
\newblock Wearable technologies: The role of usefulness and visibility in smartwatch adoption.
\newblock {\em Computers in Human Behavior}, 65:276--284, 2016.

\end{thebibliography}
\bibliographystyle{unsrt}

\end{document}